\documentclass[aps,prl,reprint,groupaddress]{revtex4-2}
\usepackage{inputenc,amsfonts,amsmath,amsthm,amssymb,amsmath,amscd,graphicx,mathrsfs,braket,hyperref,xcolor}

\renewcommand\vec{\boldsymbol}
\def\r{{\vec{r}}}

\def\p{{\vec{p}}}
\def\v{{\vec{v}}}
\def\q{{\vec{q}}}
\def\A{{\vec{A}}}

\begin{document}
\title{Moir\'e Landau fans and magic zeros}
\author{Nisarga Paul}
\thanks{These authors contributed equally}
\author{Philip J.D. Crowley}
\thanks{These authors contributed equally}
\author{Trithep Devakul}
\thanks{These authors contributed equally}
\author{Liang Fu} 
\affiliation{Department of Physics, Massachusetts Institute of Technology, Cambridge, MA,
USA}

\begin{abstract}
We study the energy spectrum of moir\'e systems under a uniform magnetic field. The superlattice potential generally broadens Landau levels into Chern bands with finite bandwidth. However, we find that these Chern bands become flat at a discrete set of magnetic fields which we dub ``magic zeros''. The flat band subspace is generally different from the Landau level subspace in the absence of the moir\'e superlattice. By developing a semiclassical quantization method and taking account of superlattice induced Bragg reflection, we prove that magic zeros arise from the simultaneous quantization of two distinct $k$-space orbits. For instance, we show the chiral model of TBG has flat bands at special fields for \textit{any} twist angle in the $n$th Landau level for $|n|>1$. The flat bands at magic zeros provide a new setting for exploring crystalline fractional quantum Hall physics.   
\end{abstract}
\maketitle

The advent of moir\'e materials has opened a new regime for the study of Bloch electrons under a magnetic field \cite{Thouless1982Aug, Harper1955Oct,Hofstadter1976Sep,wilkinson1984critical,Bistritzer2011Jul2}. 
%In natural solids with an angstrom-scale lattice constant,
%the quantization of $k$-space orbits results in a discrete set of Landau levels that are infinitely degenerate. 
Moir\'e materials feature a superlattice period that is much larger than the atomic spacing and can be comparable to the magnetic length at $B=1$T ($26$nm).   
Moreover, the superlattice potential that creates mini-bands 
is weak and slowly varying. % on the atomic scale. %, and the magnitude of its spatial variation is small.   
%For example,  %the superlattice potential is less than 100meV in twisted bilayer graphene, 
%the superlattice potential in semiconductor transition metal dichalcogenide (TMD) heterostructures with $K$-valley moir\'e bands is only on the order of 10meV.  
As a result of both features, the interplay between Landau quantization and superlattice potential can give rise to a complex energy spectrum and novel quantum phenomena not found in ordinary solids \cite{Li2021May,Forsythe2018Jul,PhysRevLett.121.026806,Huber2020Nov,Hunt2013Jun,Dean2013May,Okada2012Oct, Spanton2018Apr,Andrews2020Jun}.   
%Much focus has been given to the fractal-like spectrum, Hofstadter's butterfly, in the tight-binding regime \cite{}. On the other hand, distinctive features of Landau spectrum in the physically realistic regime of {\it weak superlattice potential} %, which is applicable to various graphene and semiconductor moir\'e materials \cite{Li2021May,Forsythe2018Jul},  
%twisted bilayer WSe$_2$, graphene/boron nitride superlattice and graphene subject to a  periodic electrostatic potential created by a patterned dielectric, 
%have been less explored \cite{Okada2012Oct,Hunt2013Jun,Dean2013May,Huber2020Nov,Lee2018Jul,Spanton2018Apr}. 
%relevant to a wide range of moir\'e materials        
%such as the Hofstadter's butterfly spectra, as observed in graphene superlattices. 

In this work, we study the energy spectrum of two-dimensional moir\'e materials under a magnetic field $B$. 
%A wide range of magnetic fields $B$ is considered: the cyclotron energy in the absence of moir\'e superlattice  can be smaller or larger than the superlattice potential. 
Our work mainly focuses on %---but is not limited to---
the case of a superlattice potential  not too strong relative to bandwidth such that the corresponding moir\'e bands can be treated by nearly free electron approximation. The energy spectrum as a function of magnetic field displays three distinct regimes. 
At very small magnetic fields, a set of Landau levels (LLs) arise from the standard semiclassical quantization  \cite{Onsager1952Sep} of cylotron orbits  that follow the constant energy contour of moir\'e bands. In the opposite limit of very large fields, a different set of LLs which come from ``free'' electrons without moir\'e effects are recovered. In the wide range of intermediate magnetic fields, the competition between magnetic breakdown and superlattice induced Bragg reflection at the mini Brillouin zone boundary leads to a new type of energy spectrum with remarkable universal features, which is the main finding of this work.  
%This regime is realized in TMD moir\'e heterostructures where the mini-bandwidth is . 
  %We also consider a wide range of magnetic fields with the flux through the moir\'e unit cell $\Phi$      
%This condition is easily satisfied in semiconductor transition metal dichacogenide based moir\'e heterostructures, 
%with a period ranging from sub-40 nm to xxx 
%\cite{Li2021May,Forsythe2018Jul}. 
%In all these systems, magnetotransport measurements observed intricate patterns in the Landau level spectra. 
%In the last example, the electrostatic potential is spatially varying in one direction only,  and in such case, Hofstadter's butterfly is absent.  

We develop a general method to calculate the Landau spectrum on the moir\'e superlattice. % and identify remarkable universal features. 
We show that at intermediate magnetic fields, the LLs of free electrons are generally broadened by Bragg scattering off the moir\'e superlattice, or in a complementary way, the LLs of Bloch electrons are broadened by magnetic breakdown near the mini Brillouin zone boundary.  
Remarkably, flat bands are found at a discrete set of magnetic fields, which we dub ``magic zeros". Plotted in the $(B, \mu)$ plane where $\mu$ is the chemical potential, each zero occurs at the intersection of two fictitious LL fans, corresponding to the simultaneous quantization of two distinct $k$-space orbits. The corresponding density of states divergence predicted by our theory directly manifests as a peak in the compressibility $dn/d\mu$. Alternatively, LL widths can be measured directly by STM \cite{Okada2012Oct} and inferred from inter-LL optical transitions \cite{Ju2017Nov}. One application is chiral twisted bilayer graphene (TBG), which we find has magic zeros in LLs for $|n|>1$ at \textit{all} twist angles and not just the discrete set of magic angles \cite{Tarnopolsky2019Mar}.

Importantly, we show the existence of these flat bands is robust and not limited to the particular known case of Schr\"odinger or Dirac LLs perturbed by a weak potential $V_0 \ll \omega_c$ (with $\omega_c$ the cyclotron frequency)\cite{Pfannkuche1992Nov}. In contrast, our theory of magic zeros is non-perturbative in $V_0/\omega_c$ and applicable to generic energy dispersions. We show the flat band at a magic zero spans a Hilbert space that is generally distinct from the  LL subspace of free electrons. The physics of flat bands at magic zeros contrasts and complements the broadening and Hofstadter-type splitting of LLs at generic $B$ fields. 
%As we prove this within a semiclassical model %(which does not depend on a particular form of the energy dispersion, meaning that 
%We show that the presence of magic zeros is a universal feature of the Landau spectrum of Bloch electrons in both 1D and 2D periodic potential. 
%As a result, the semiclassical model provides a unifying picture for the range of systems in 1-5. 
     
\par 

%Our calculation of moir\'e Landau spectrum and identification of magic zeros are based on a combination of methods. 
%We develop a general method for calculating the moir\'e Landau spectrum, which is based on semiclassical quantization of $k$-space orbits in the presence of Bragg scattering off the moir\'e potential. Importantly, our method is applicable to not only moir\'e semiconductors (such as WSe$_2$/WS$_2$), but also moir\'e metals such as twisted bilayer NbSe$_2$, which has a large Fermi surface and cannot be described by the  continuum model approach. While our semiclassical method is not, strictly speaking, valid for the first few Landau bands, for both one-dimensional and two-dimensional periodic potentials, by numerically solving the energy spectrum over a wide range of potential strengths we find magic zeros in all Landau bands, except the lowest one that develops out of the $n=0$ Landau level. Our analytical and numerical findings show that the presence of magic zeros is a robust feature of the energy spectrum of Bloch electrons under a magnetic field.  

%The study of the magnetic spectrum of a system such as TBG typically involves starting from a continuum model. However, many interesting systems such as moir\'e metals or highly-doped semiconductors have a large Fermi surface and no such continuum approximation is useful. In such cases, it is still worth asking whether some universal features of the Landau level spectrum emerge due to the moir\'e periodicity. 

%quantum Hall effect in two-dimensional semiconductor structures.      

\begin{figure}
\includegraphics[width=1.0\columnwidth]{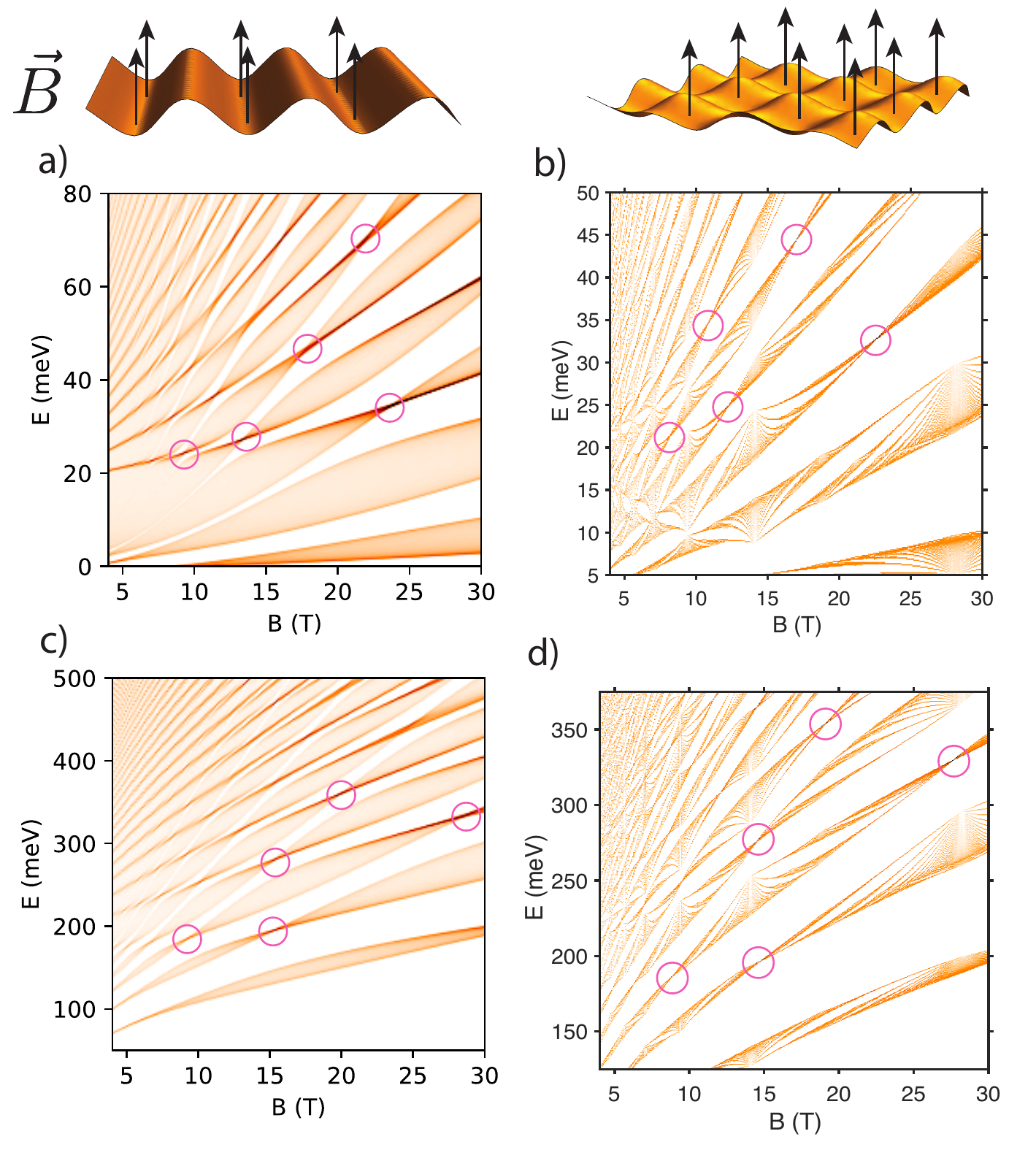}
\caption{DOS in a periodic potential and magnetic field for Schr\"odinger (a,c) and Dirac (b,d) electrons. A few prominent ``magic zeros" are circled, robust features where the bandwidth vanishes. (a,b) Exact diagonalization for a 1D periodic potential. (c,d) Perturbative results for a 2D six-fold symmetric potential. Parameters: (a) $V_0 = 15$ meV, $m^* = 0.2 m_e$, (b) $V_0 = 70$ meV, $v=10^6$ m/s, (c) $V_0=4$ meV, $m^* = 0.2m_e$, (d) $V_0=25$ meV, $v=10^6$ m/s. Period $a=13$ nm.}\label{fig:ED}
\end{figure}

\par 
We consider a two-dimensional Bloch electron in a uniform magnetic field:
\begin{equation}\label{eq:H}
    H = H_0(\p-e\A)+ V(\r)
\end{equation}
where $H_0(\p)$ denotes the energy dispersion in the absence of moir\'e superlattice 
and $\A$ is the vector potential.  
\begin{equation}\label{eq:V(r)}
        V(\r) = \sum_{\q} V(\q) e^{i\q \cdot \r} + c.c. 
\end{equation}
denotes a periodic moir\'e potential ($\hbar = 1$). As the superlattice potential in moir\'e materials  is slowly varying, $V(\r)$ is well approximated as a sum of a few lowest leading harmonics.

Depending on the form of $H_0$ and $V$, $H$ describes a wide variety of moir\'e materials. 
In the case of semiconductor transition metal dichalcogenide (TMD) heterostructures such as WSe$_2$/WS$_2$, 
$H_0(\p) = p^2/2m$ where $m$ is the effective mass near the band edge of TMD monolayer, and the triangular symmetric potential $V(\r)$ is composed of three Fourier components of equal magnitude at wavevectors related by symmetry \cite{Jin2019Mar}.   
In the case of graphene on a one-dimensional patterned dielectric superlattice, $H_0(\p) = v \p \cdot \sigma$ is the Dirac Hamiltonian of graphene, and $V(\r) =  V_0 \cos(q x)$ involves a single wavevector only \cite{Li2021May}.  
In both cases, the periodic potential  $V(\r)$ results in  mini-bands, as manifested in resistive peaks at commensurate densities. Under a magnetic field, transport measurements  observed complex patterns in the LL spectra.

The first indication of magic zeros can be found in the regime where the superlattice potential strength is smaller than the cyclotron energy $\omega_c$ of free electrons \cite{Pfannkuche1992Nov}. In this perturbative regime, $V(\r)$ lifts the infinite degeneracy within a LL. The  projection of periodic potential into the $n$th LL of Schr\"odinger electrons can then be written \cite{Wilkinson1987May,Pfannkuche1992Nov} (choosing symmetric gauge $\A = \frac12 \vec B \times \r$)
\begin{equation}\label{eq:H1n}
V^{\text{eff}}_n=\sum_\q V(\q) e^{-q^2l_B^2/4}L_n(q^2l_B^2/2) e^{i\q \cdot (-\tilde{\pi}_y,\tilde{\pi}_x)l_B^2}
\end{equation}
where $l_B = 1/\sqrt{eB}$ is the magnetic length, $\vec{\tilde\pi}= \p + e\A$, and $L_n$ is the $n$th Laguerre polynomial. 
Note $[\tilde{\pi}_x,\tilde{\pi}_y]=-ieB$. Notably, when all wavevectors are of equal magnitude $q$, the LL projected potential in Eq. \eqref{eq:H1n} vanishes at $n$ values of $ql_B$ due to the Laguerre polynomial zeros, leading to a flat Chern band despite the presence of periodic potential. %The $n$th level exhibits $n$ magic zeros. 
In the case of Dirac electrons, the $n$th LL wavefunction is a two-component spinor and the projected potential is given by \cite{Huber2021Jun}:  $\tilde{V}^{\text{eff}}_n = (V^{\text{eff}}_{|n|} + V^{\text{eff}}_{|n|-1})/2$ for $n\neq 0$. %and $\tilde{V}^{\text{eff}}_0 = V^{\text{eff}}_{0}$. 
Zeros occur in this case as well. Magic zeros for the first few LLs are listed in Table \ref{table}, and the perturbative spectrum for a six-fold symmetric potential is shown in Fig. \ref{fig:ED}c-d.

\begin{table}[t!]%The best place to locate the table environment is directly after its first reference in text
\begin{ruledtabular}
\begin{tabular}{l|l|l}
\textrm{$|n|$}&
$ql_B$ \textrm{(Schr\"odinger case)}&
$ql_B$ \textrm{(Dirac case)}\\
\colrule
$1$ &  $\sqrt{2}$ & 2 \\
$2$ & 1.08, 2.61 & 1.24, 3.24 \\
$3$ & 0.91, 2.14, 3.55 & 0.99, 2.36, 4.18 \\
$4$ &0.80, 1.87, 3.01, 4.34& 0.86, 2, 3.26, 4.96\\
$5$ &0.73, 1.68, 2.68, 3.77, 5.03 & 0.76, 1.77, 2.84, 4.03, 5.65\\
\end{tabular}
\end{ruledtabular}
\caption{Values of $ql_B$ for which the $n$th LL has zero bandwidth for weak potential, where $q$ is the potential wavevector and $l_B$ the magnetic length, for Schr\"odinger and Dirac electrons. The $n$th level exhibits $|n|$ magic zeros.}\label{table}
\end{table}

\par 

Remarkably, we find that the magic zeros persist beyond the perturbative regime, as indicated by the exact diagonalization (ED) of the energy spectrum of Eq. \eqref{eq:H} in Fig. \ref{fig:ED}a-b for the case of a potential $V_0\cos(qx)$. At the density of states (DOS) peaks shown, the bandwidth is zero within numerical accuracy, even in the regime $V_{0}/\omega_c\sim 3$. This result is truly all-orders in $V_0/\omega_c$, as indicated by the following: (i)  magic zeros deviate from the Laguerre polynomial zeros and (ii) the wavefunction at the zeros differs from the LL wavefunction at $V(\r)=0$ (see SM).   
In order to uncover the origin of these zeros, we develop a semiclassical approach which places no restrictions on $V_{0}/\omega_c$. Moreover, this approach does not rely on a specific form of energy dispersion $H_0(\p)$,  %While the ED plots were for particular energy dispersions, 
and thus is applicable to a wider range of systems, such as bilayer graphene with trigonal warping. The starting point for the semiclassics is to consider a Bloch wavepacket whose position and momentum are governed by the equations
\begin{equation}\label{eq:pdot}
    \dot\p = -e  \dot \r \times \vec B,\quad 
    \dot \r  = \mathbf{\nabla} E(\p),
\end{equation}
where $E(\p)$ is the energy dispersion \textit{including} the effect of the periodic potential. 

When the potential $V(\r)$ is absent, electrons at an energy $\varepsilon$ follow the original Fermi surface $H_0(\p) = \varepsilon$. When the potential is strong, electrons follow the reconstructed Fermi surface $E(\p)=\varepsilon$ where $\p$ lies in the mini Brillouin zone. In both cases, semiclassical quantization predicts infinitely degenerate LLs whenever the real-space orbits, which are simply $\p$-space orbits rotated by $\pi/2$ and scaled by $1/B$, enclose integer flux \cite{Onsager1952Sep}.\par 

In between these two limits, magnetic breakdown \cite{Blount1962Jun,Cohen1961Sep,Reitz1964Jan} broadens the LLs. Let us consider an intersection of two original Fermi surfaces at the first Bragg plane in the repeated-zone scheme. In a magnetic field, there are two incoming and two outgoing electron wavepackets. Thus we may treat the intersection as a two-level Landau-Zener system with a scattering matrix $U$. When $B$ is sufficiently small, electrons follow on the reconstructed Fermi surface and occasionally break through, and $U$ is mainly diagonal. When $B$ is large, electrons follow the original Fermi surface and occasionally Bragg scatter, so $U$ is mainly off-diagonal. 

In general, the scattering matrix takes the form
\begin{equation}\label{eq:U}
 U=   \begin{pmatrix}
    \sqrt{1-P}e^{-i\widetilde \varphi_S} & -\sqrt{P}\\
    \sqrt{P} & \sqrt{1-P}e^{i\widetilde \varphi_S}
    \end{pmatrix}
\end{equation}
where the magnetic breakdown probability is
\begin{equation}\label{eq:PLZ}
    P = e^{-2\pi/\delta}, \quad \delta = 16 eB v_1v_2 \sin\alpha/E_{\text{gap}}^2,
\end{equation}
$\v_1$ and $\v_2$ are incoming electron velocities which differ by an angle $\alpha$, $E_\text{gap}=2V_0$ is the band gap at the Bragg plane, and $\widetilde \varphi_S = \varphi_S-\pi/2$ with $\varphi_S = \pi/4-(\ln \delta +1)/\delta+\arg \Gamma(1-i/\delta)$, the so-called Stokes phase \cite{Blount1962Jun,Shevchenko2010Jul} (see SM). We note $e^{-i\widetilde\varphi_S}$ only depends weakly on $\delta$, interpolating between $i$ and $i^{1/2}$. Eqs. \eqref{eq:U} and \eqref{eq:PLZ} are derived using the nearly free electron approximation assuming that the effect of the potential on the band structure is only significant near Brillouin zone boundary. Note that $P$ goes to zero quickly at low fields and approaches $1$ at high fields. 

In the case of parabolic bands, $\delta$ reduces to $8\varepsilon\omega_c\sin\alpha/V_0^2$. For bilayer TMDs with $\varepsilon\sim 20$ to $ 40$ meV, $q\sim k_F$ ($k_F$ is the Fermi wavevector), and $\omega_c\sim 2$ meV at 10 T, $P \sim 0.1$ to $0.3$. For graphene in a 1D potential \cite{Li2021May}, taking $V_0 a/v_F \sim 1$ to $10$, $a\sim 50$ nm, $q\sim k_F$, and $B \sim 10$ T gives $P\sim 0.01$ to $0.95$. Evidently, realistic values of $P$ in moir\'e materials require that the effects of magnetic breakdown are properly taken into account. \par

\begin{figure}
\includegraphics[width=1.0\linewidth]{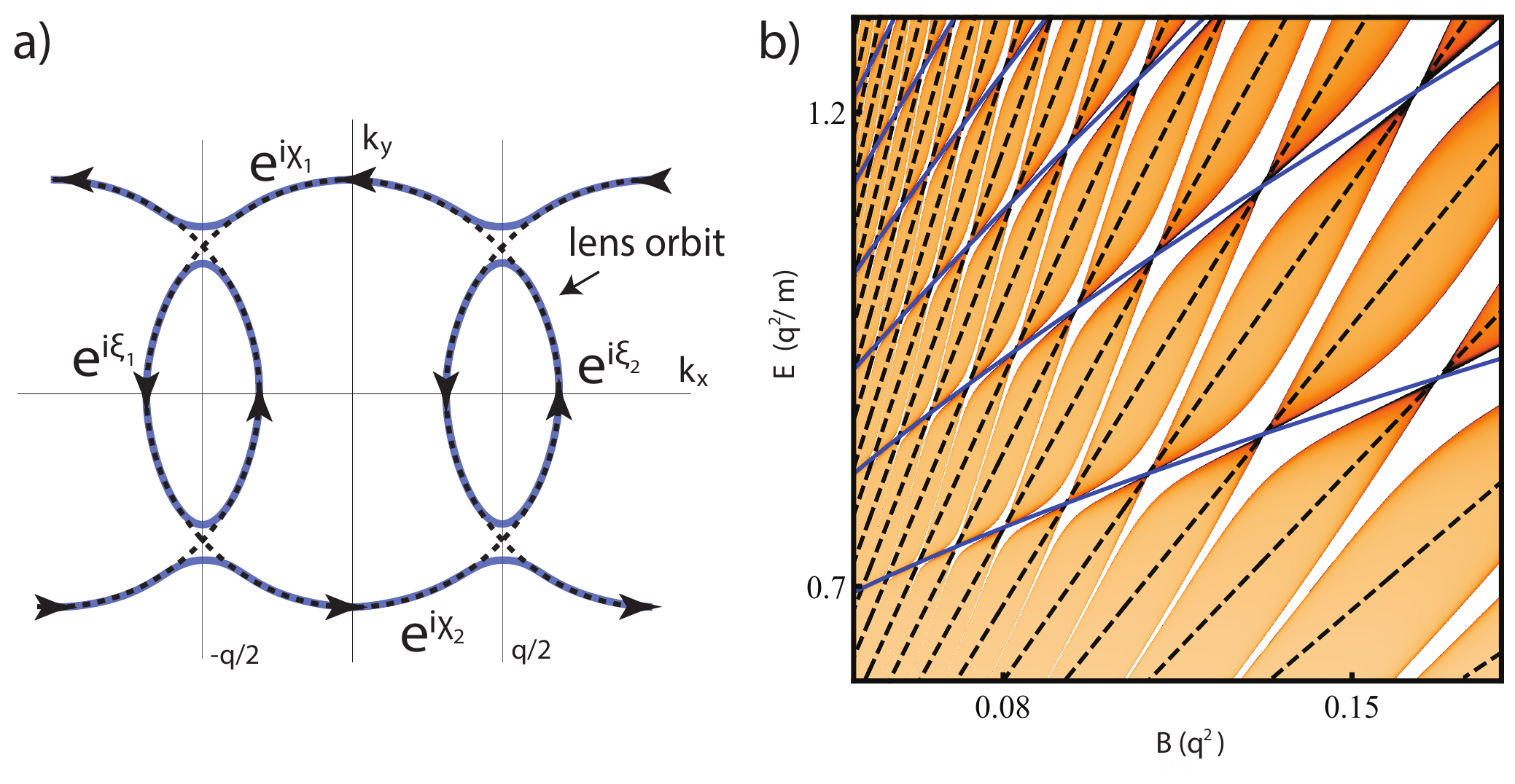}
\caption{(a) Fermi surface in the repeated zone scheme in the presence of a 1D potential $V_0\cos(qx)$. The network model (dashed lines) involves scattering at intersections and phases picked up on links. (b) Intersecting Landau fans due to the original orbit (dashed lines) and lens orbit (blue), and semiclassical DOS (density plot) with $V_0 =0.4, q=2,m=1$.}\label{fig:schematic}
\end{figure}

To properly account for magnetic breakdown, we consider a network model comprised of the original Fermi surfaces in the repeated zone scheme where wavepacket motion away from the intersections is free electron-like while scattering at the intersections is given by the Landau-Zener unitary $U$. Let us first consider networks in which neighboring Fermi surface intersect at two points as in Fig. \ref{fig:schematic}a. This is similar in spirit to models considered by Pippard \cite{Pippard1962Oct,Pippard1964Jul}; we refer to \cite{PhysRevResearch.2.033271,Chambers1965Oct,Chambers1966Jul} for other examples of network model constructions. We refer to the original Fermi surfaces defined by $H_0(\p) = \varepsilon$ as the ``original orbit" and their intersection as the ``lens orbit". 

To understand the magic zero condition in the semiclassical approach, it is instructive to consider the scattering matrix across a lens orbit, which is given by 
\begin{subequations}\label{eq:W}
\begin{align}
    W&= \frac{1}{(1-P)e^{i(\xi_1+\xi_2+2\widetilde\varphi_S)}-1} \begin{pmatrix}
    Pe^{i\xi_1} & \kappa \\
    \kappa & Pe^{i\xi_2}\\
    \end{pmatrix}\\
    \kappa &= e^{-i\widetilde\varphi_S}\sqrt{1-P}(e^{i(\xi_1+\xi_2+2\widetilde\varphi_S)}-1),
\end{align}
\end{subequations}
where $\xi_1,\xi_2$ are the phases acquired along the links of the lens orbit. $W$ describes scattering between incoming and outgoing states across the lens orbit. When $W$ is diagonal, incoming states scatter into outgoing states in the same zone.  
\par 
We note that when $\xi_1+\xi_2 + 2\widetilde \varphi_S$ is an integer multiple of $2\pi$, $W$ is diagonal, indicating the decoupling of neighboring orbits in the network.  This is reminiscent of constructive interference in a Fabry-P\'erot optical cavity \cite{PhysRevLett.121.026806,Drienovsky2014Mar}, where the lens orbit plays the role of the cavity. The decoupled orbits are valid eigenstates when the phase around the original orbit is an integer multiple of $2\pi$. Under these conditions, the network model supports an extensive set of states which are localized and dispersionless, i.e. a flat band. 
\par
In brief, the flat band conditions, in terms of the phases shown in Fig. \ref{fig:schematic}a, are
$ \sum \xi_i + 2\widetilde \varphi_S \in 2\pi \mathbb{Z}$ and $\sum (\xi_i+ \chi_i) \in 2\pi \mathbb{Z}$. The phases satisfy:
\begin{equation}\label{eq:phases}
    \sum_\mathrm{orig.} \xi_i + \chi_i  = l_B^2 S_0+2\pi \gamma, \quad\sum_\mathrm{lens}\xi_i  = l_B^2 S_1+2\pi \gamma
\end{equation}
where $S_0, S_1$ are the $\p$-space areas of the original and lens orbits, respectively. We have added the topological Maslov contribution $\gamma=1/2 -\varphi_{\text{Berry}}/2\pi$ which is customary in semiclassical treatments for closed orbits deformable to a circle \cite{Xiao2010Jul,Keller1958Jun,Mikitik1999Mar,Fuchs2010Oct}, with Schr\"odinger and Dirac electrons having $\gamma=1/2$ and $0$ respectively. $\varphi_\text{Berry}$ is the Berry phase along the orbit. \par 

Combining the above conditions, we find that bandwidth zeros occur at the intersection of the two Landau fans given by 
\begin{subequations}\label{eq:fans}
\begin{align}
    l_B^2 S_0 &= 2\pi (n+\gamma)\label{eq:fan0}\\
    l_B^2 S_1 &= 2\pi (m+ \gamma - \widetilde \varphi_S/\pi)\label{eq:fan1}
\end{align}
\end{subequations}
for suitable integers $m,n$. These equations stipulate that both the original and lens orbits enclose integer flux, up to the Stokes phase and Maslov correction. 
%We note that $\widetilde \varphi_S/\pi \mod 1$ varies over the range $[3/4,1/2)$. \par 
The magic zero conditions, i.e. Eq. \eqref{eq:fans}, can be roughly thought of as Bohr-Sommerfeld quantization conditions for both the original and lens orbits. In general, a sufficient condition for magic zeros is a Fermi contour with only \textit{two} relevant independent semiclassical electron orbits (with other orbit areas integer linear combinations of these). For circular Fermi surfaces, these two areas are
\begin{equation}\label{eq:S0S1}
    S_0 = \pi k_F^2,\quad
    S_1 = 2k_F^2 (\cos^{-1}x - x\sqrt{1-x^2}),
\end{equation}
where $x= q/2k_F$. For this case the intersecting Landau fans are shown in Fig. \ref{fig:schematic}b. \par

In the large $n$ and weak potential limit, the semiclassical and perturbative approaches are expected to agree. Eq. \eqref{eq:S0S1} and Eq. \eqref{eq:fan0} give $k_F^2 = 2(n+\gamma)/l_B^2$. Applying this to Eqs. \eqref{eq:fan1} and \eqref{eq:S0S1} and noting $\widetilde\varphi_S/\pi \to -1/4$ at weak potential, the values of $ql_B$ at which the $n$th LL has a magic zero are given by 
\begin{equation}\label{eq:qlb}
    ql_B = \frac{\pi (n-m-1/4)}{\sqrt{2n}} + O(n^{-1})
\end{equation}
for integers $m,n$. In the perturbative regime, Eq. \eqref{eq:H1n} implies that these are the zeros of $L_n(q^2l_B^2/2)$. Indeed, by applying the large $n$ formula \cite{szego1975}
\begin{equation}\label{eq:limit}
    e^{-\frac{q^2l_B^2}{4}}L_n(q^2l_B^2/2) =  \frac{ \cos(\sqrt{2n}ql_B-\frac{\pi}{4})}{\sqrt{\pi q l_B \sqrt{n/2}}} + O(n^{-\frac34}),
\end{equation}
we see that these magic zero conditions derived independently are identical. We remark that the phenomenon of Weiss oscillations \cite{Gerhardts1989Mar,Zhang1990Jun,Steda1990Jun,Pfannkuche1992Nov,PhysRevLett.121.026806,Beenakker1989Apr}---superlattice induced magnetoresistance oscillations---is naturally captured by the semiclassical approach in this regime (see SM).  \par 

% As was to be expected, the bandwidth zero conditions we found reproduce the more familiar commensurability condition in the regime of overlap: at large $n$ in the perturbative limit for a circular Fermi surfaces. However, Eq. \eqref{eq:2Rc} cannot be not extended much beyond this regime. For instance, it neglects higher-order perturbative corrections and lacks a precise definition for $R_c$ for non-circular Fermi surfaces. In contrast, the conditions we provided, phrased as Bohr-Sommerfeld-type quantization rules, represent an all-orders extension of the commensurability condition far beyond its domain of validity. 

\begin{figure}[t]
    \centering
    \includegraphics[width=1.0\columnwidth]{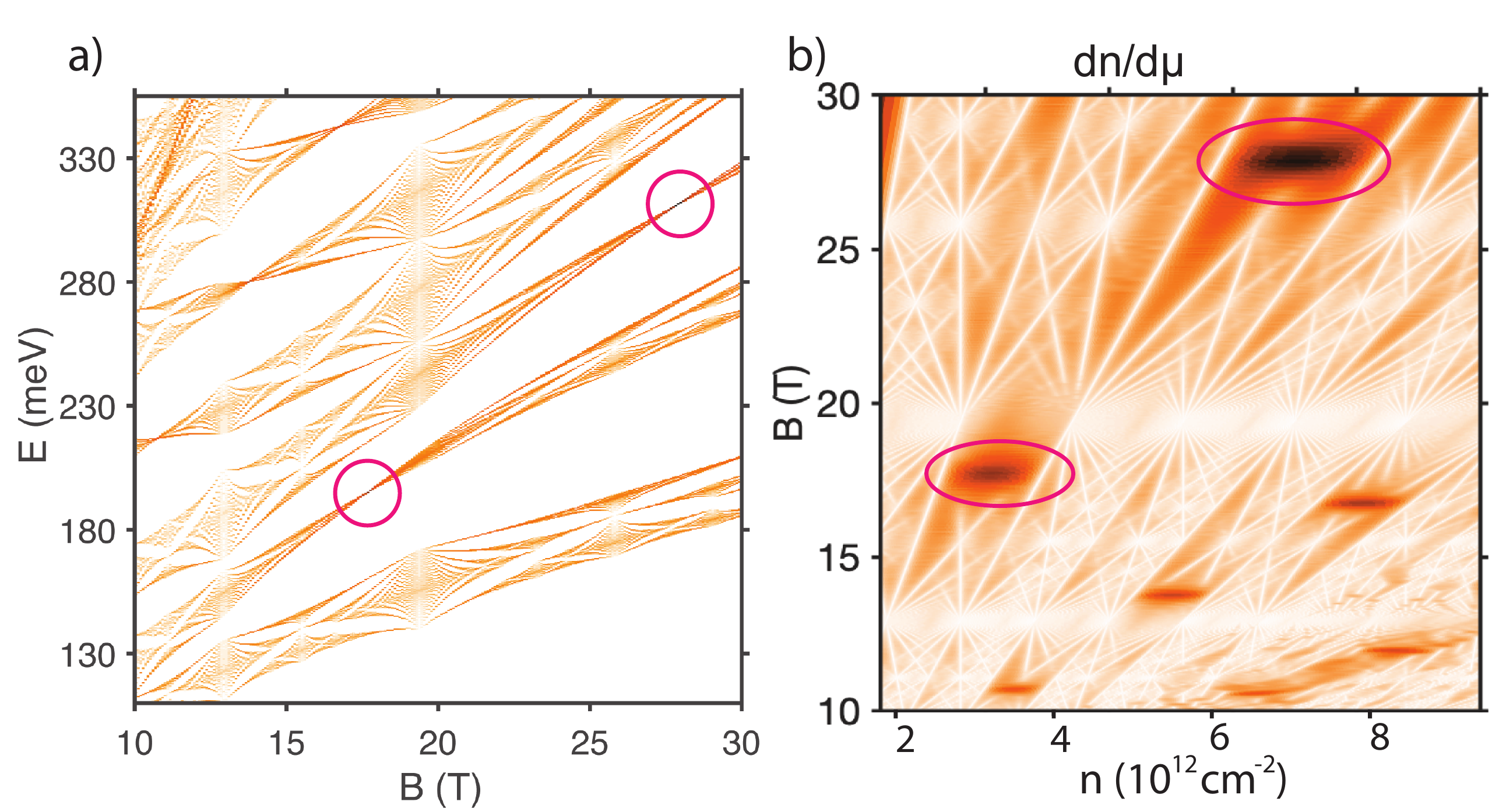}
    \caption{(a) DOS for chiral limit TBG with $V_0 = 30$ meV, $v = 10^6$ m/s at $\theta = 1.1^\circ$. Zeros exist at any twist angle. (b) Noninteracting compressibility as a function of density and $B$ at $T=0.2$ meV. Magic zeros are dark features of high compressiblity occurring over a finite range of $n$.}
    \label{fig:compressibility}
\end{figure}

So far we have only discussed magic zeros, but the network model also allows us to calculate the band dispersion at generic fields using a transfer matrix approach. For the network in Fig. \ref{fig:schematic} due to a 1D potential (we defer discussion of the 2D case), the transfer matrix eigenvalues $e^{\pm i\theta}$ satisfy the relation
\begin{equation}
    \cos\theta = \frac{\sin(\xi+\chi) + (1-P)\sin(\xi-\chi+2\widetilde \varphi_S)}{2\sqrt{1-P}\sin(\xi+\widetilde\varphi_S)}
\end{equation}
where a gauge choice such that $\xi_1=\xi_2=\xi$ and $\chi_1=\chi_2=\chi$ has been made. The resulting semiclassical spectrum is shown in Fig. \ref{fig:schematic}b. We show the quantitative agreement with ED in the SM. The LLs alternately broaden and pinch off at magic zeros, and the corresponding DOS divergences directly manifest as peaks in compressibility $dn/d\mu$  (see e.g. Fig. \ref{fig:compressibility}b and the SM). Importantly, we have placed no restrictions on $V_0/\omega_c$, so our results are all-orders in conventional perturbation theory. Moreover, our treatment did not depend on the precise energy dispersion, and a different dispersion would only alter geometric details such as Fermi surface areas and link phases. The existence of magic zeros, which is our main focus, is robust to all these details.  \par 

So far we have assumed the Fermi surface intersects only a single pair of Bragg planes at $\pm q \hat{\vec{x}} /2$.  In moir\'e materials made of highly doped semiconductors or metals, however, $V(\vec{r})$ may have a small wavevector compared to the size of the Fermi surface, resulting in a network many overlaps. An important simplification is that the gaps at the intersections form a distinct hierarchy with $E^{n^\text{th}}_\text{gap}\sim V_0^n/(v_Fk_F)^{n-1}$ at an $n$th order Bragg plane. Therefore $P$ has a double-exponential dependence on $n$ and only a few crossings are active, with the rest completely avoided ($P=0$) or trivial ($P=1$). The simplest scenario is when only the intersections at the first-order Bragg plane are active. From Eq. \eqref{eq:PLZ}, for the parabolic case this requires
$
V_0^4/(v_Fk_F)^3 \ll \omega_c \sin\alpha_2, V_0^2/v_Fk_F \not\ll \omega_c \sin\alpha_1
$
where $\alpha_j$ is the intersection angle at the $j$th Bragg plane (and $\alpha_1\approx \alpha_2$ when intersections are close together). In this ``first-order regime" \cite{Spurrier2019Aug}, the network model maps exactly back onto the simplest case of a single intersection, Fig. \ref{fig:schematic}a. Therefore the DOS plot is the same as before, albeit with a slightly restricted regime of validity. \par 

Let us discuss the extension to 2D potentials, such as a triangular lattice potential. Strictly speaking, the network model approach is only valid when $\varphi$, the number of flux quanta per real-space unit cell, is a rational number $p/q$; then the network unit cell is enlarged by a factor of $p$ and each LL contains $p$ subbands (for coprime $p,q$)~\cite{Pippard1964Jul}. However, if the enlarged unit cell consists of only original and lens orbits, they decouple when $W$ becomes diagonal, and the flux at a magic zero can be approximated arbitrarily well by a rational $\varphi$. Thus magic zeros arise in this case with identical conditions. When the network topology has overlapping lens orbits, we conjecture that magic zeros still arise (see SM). For magic zeros to arise, the lens orbits of the network model must have equal areas, which requires wavevectors of equal magnitude. This is consistent with the perturbative result of Eq. \eqref{eq:H1n}. \par
We underscore a few necessary conditions for magic zeros which are frequently satisfied in moiré materials: (1) a superlattice potential consisting of equal-magnitude wavevectors must be present, (2) the Fermi surface must be invariant under the rotational symmetry of the potential, and (3) the potential is not so strong as to significantly restructure the bandstructure. Perturbations which would broaden magic zeros include higher order harmonics, strain, or anisotropic dispersions.  \par 
A natural question is whether TBG \cite{Bistritzer2011Jul,Cao2018Apr} exhibits magic zeros. While our theory based on a scalar moiré potential does not apply directly, we find magic zeros exist in the chiral limit \cite{Tarnopolsky2019Mar} at \textit{any} twist angle; we plot the noninteracting perturbative DOS and compressibility in Fig. \ref{fig:compressibility} and include details in the SM.
\par 
The flat Chern band at magic zeros provides an ideal setting for realizing fractional quantum Hall (FQH) states and other novel states at fractional fillings. For instance, %in the case of a one-dimensional periodic potential, 
the increased bandwidth away from magic zeros weakens the Laughlin state and may induce a transition into metallic states or electron crystals. %Moreover, the projected Coulomb interaction will generally depend on the periodic potential, possibly leading to various anisotropic FQH states. 
We leave these directions to future work. \par 
\begin{acknowledgments}
\textit{Acknowledgments--} We thank Ray Ashoori and Long Ju for helpful discussions. This work is supported by a Simons Investigator Award from the Simons Foundation and  the STC Center for Integrated Quantum Materials (CIQM) under NSF award no. ECCS-2025158. LF is partly supported by the David and Lucile Packard Foundation.
\end{acknowledgments}
\bibliographystyle{apsrev4-1}
\bibliography{bib}

\end{document}

\title{Supplemental Material}
\author{Nisarga Paul}
\thanks{These authors contributed equally}
\author{Philip J.D. Crowley}
\thanks{These authors contributed equally}
\author{Trithep Devakul}
\thanks{These authors contributed equally}
\author{Liang Fu} 
\affiliation{Department of Physics, Massachusetts Institute of Technology, Cambridge, MA,
USA}

\maketitle

\section{Wavefunctions at magic zeros}

\begin{figure}
    \centering
    \includegraphics[width=1.0\columnwidth]{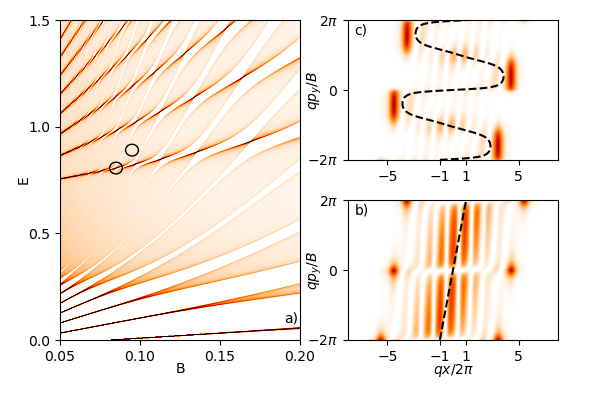}
    \caption{(a) Density of states from exact diagonalization for parabolic dispersion $p^2/2m$ and periodic potential $V_0 \cos(qx)$ with $V_0=0.4, q=2$ and $m=1$. (b) Each horizontal line cut shows the Landau gauge wavefunction profile at a given momentum $p_y$ away from a magic zero and (c) at a magic zero. The dashed line shows $\braket{x}$.}
    \label{fig:wvfn}
\end{figure}

For a dispersion $p^2/2m$ with Landau gauge $\A= Bx \mathbf{y}$, Landau level wavefunctions take the form 
\begin{equation}\label{eq:wvfn}
    \psi_{n,p_y}(x,y) \sim e^{ip_y y}H_n(x-p_yl_B^2)e^{-(x-p_yl_B^2)^2/2l_B^2}
\end{equation}
and have a mean position $\braket{x}= p_yl_B^2$. In Fig. \ref{fig:wvfn}, we plot the wavefunctions after including a periodic potential $V_0 \cos qx$. At generic fields, $\braket{x}$ has a nonlinear dependence on $p_y$ while at magic zeros, a linear dependence $\braket{x} \sim p_y$ is recovered. However, the wavefunctions at a magic zero span a subspace distinct from the Landau level subspace spanned by free electron wavefunctions, which is clear because Fig. \ref{fig:wvfn}b corresponds to a functional form distinct from Eq. \eqref{eq:wvfn}.

\section{Details of the semiclassical model}

\textit{Scattering between bands as a Landau-Zener effect. } In this section we present the mapping of a wavepacket scattering between two bands at an avoided crossing to a Landau-Zener problem, rectifying the original derivation in Ref.  \cite{Blount1962Jun}. Consider a wavepacket in $\k$-space prepared initially in a single band with dispersion $E(\k)$. In the presence of a magnetic field this wavepacket moves along the equipotentials of the dispersion. Consider introducing a periodic modulation in the extended zone scheme so we have states at all momenta $E(\k+\g)$ where $\g$ is any reciprocal lattice vector. Consider two bands $E_1(\k) = E(\k + \g_1)$, $E_2(\k) = E(\k + \g_2)$, which are (na\"ively) degenerate at the point $\k^*$, so that $E^* = E_1(\k^*) = E_2(\k^*)$. This degeneracy is ultimately split by a matrix element $V$, so that in the vicinity of $\k^*$ we have the Hamiltonian
\begin{equation}
    H = \begin{pmatrix}
    E_1 (\k) & V
    \\
    V & E_2(\k)
    \end{pmatrix}
\end{equation}
As the wave-packet passes through $\k^*$ it has some amplitude for scattering from one band to the other, an effect captured by the a $2 \times 2$ scattering matrix $U$. First we include minimal coupling $\k \to \p = \k - e \A$, and linearise the Hamiltonian about $\k^*$
\begin{equation}
    H = \begin{pmatrix}
    E^* + \v_1 \cdot \p & V
    \\
  V & E^* + \v_2 \cdot \p
    \end{pmatrix}
\end{equation}
where $\v_n = \left. \nabla_\k E_n (\k) \right|_{\k = \k^*}$. This may be written in terms of Pauli as matrices as
\begin{equation}
    H = E^* + (\v_+ \cdot \p) + V \sigma_x + (v_- \cdot \p) \sigma_z
\end{equation}
where $\v_\pm = \tfrac12 (\v_1 \pm \v_2)$. We transform to the rotating frame $\ket{\psi'} = \e^{i X t} \ket{\psi}$ with $X = E^* + \v_+ \cdot \p$ yielding the effective Hamiltonian
\begin{equation}
    H' =  V \sigma_x + (v_- \cdot \p + u t) \sigma_z
\end{equation}
where we have defined the constant $u$ via the relation
\begin{equation}
    u = i [\vec{v}_- \cdot \p,\vec{v}_+ \cdot \p]  = e \vec{B} \cdot (\v_- \times \v_+).
\end{equation}
We are now at liberty to work in the diagonal basis of $v_- \cdot \p$, so that $v_- \cdot \p$ is replaced with its eigenvalue, which may be absorbed by a redefinition of the origin of time to yield
\begin{equation}
    H' =  V \sigma_x + u t \sigma_z,
    \label{eq:H2}
\end{equation}
the textbook Landau-Zener Hamiltonian \cite{Landau1932,Zener1932Sep}. The Landau-Zener problem is characterised entirely by the dimensionless velocity
\begin{equation}
    \left|\frac{u}{V^2}\right| = \frac{2 e B v_1 v_2 \sin \alpha}{V^2}
\end{equation}
where $\alpha$ is the angle between $\v_1$ and $\v_2$. We note the recovery of the expected ratio for parabolic bands
\begin{equation}
    \left|\frac{u}{V^2}\right| = \frac{4 E_\mathrm{F} \omega_c \sin \alpha}{V^2}
\end{equation}
where $E_\mathrm{F} = m v_\mathrm{F}^2/2$ is the Fermi energy and $\omega_c = e B / m$ is the cyclotron frequency.\par 

\textit{Unitary scattering matrix $U$.} The instantaneous eigenvalues of Eq. \eqref{eq:H2} are $E_{\pm} = \pm \sqrt{\Delta^2+u^2t^2}.$ The corresponding instantaneous eigenvectors $\phi_{\pm}$ are only defined up to a (possibly time-dependent) phase. We choose a reasonable convention in which $\lim_{t\to -\infty} \phi_-= \lim_{t\to \infty} \phi_+$ and $\lim_{t\to -\infty} \phi_+ = \lim_{t\to \infty} \phi_-$. Then the appropriate unitary describing the Landau-Zener transition in this basis is \cite{Shevchenko2010Jul}
\begin{equation}
 U=   \begin{pmatrix}
    \sqrt{1-P}e^{-i\widetilde \varphi_S} & -\sqrt{P}\\
    \sqrt{P} & \sqrt{1-P}e^{i\widetilde \varphi_S}
    \end{pmatrix}
\end{equation}
where the magnetic breakdown probability is
\begin{equation}\label{eq:PLZapp}
    P = e^{-2\pi/\delta}, \quad \delta = 4 eB v_1v_2 \sin\alpha/V^2,
\end{equation}
and $\widetilde \varphi_S = \varphi_S-\pi/2$ with $\varphi_S = \pi/4-(\ln \delta +1)/\delta+\arg \Gamma(1-i/\delta)$, the so-called Stokes phase. The Stokes phase is essential in recovering zeros in the perturbative limit and was missed by other semiclassical treatments as far as we know.\par

\begin{figure}[t]
    \centering
    \includegraphics[width=1.0\linewidth]{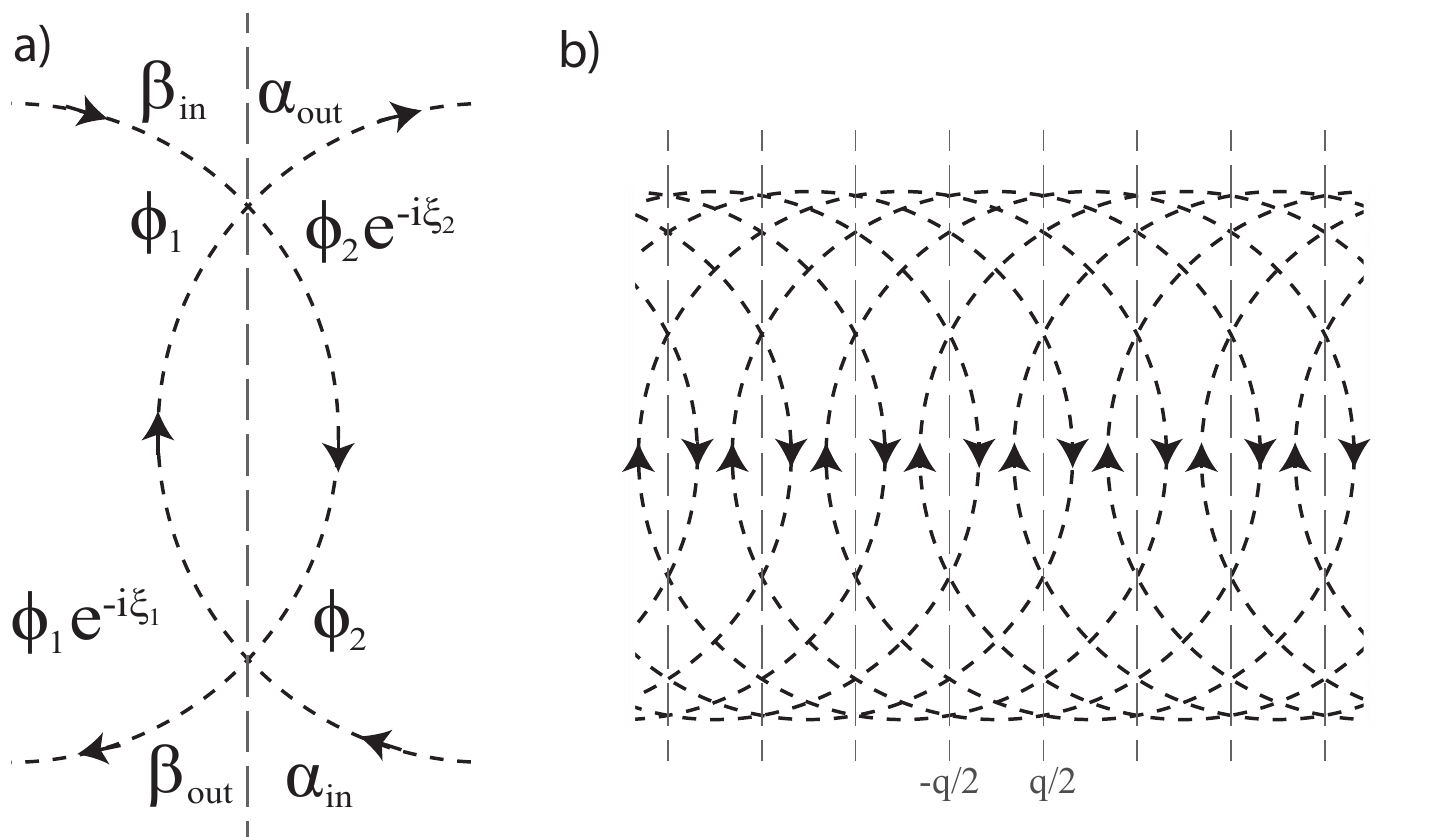}
    \caption{(a) Defining the amplitudes in the lens orbit, relevant for the derivation of the unitary $W$. (b) Network model when many Fermi surfaces overlap.}
    \label{fig:appendix1}
\end{figure}

\textit{Unitary across a lens orbit $W$. } Let us derive the unitary matrix $W$ describing scattering across a lens orbit. We define various amplitudes in Fig. \ref{fig:appendix1}a. We seek the matrix satisfying
\begin{equation}
    W\begin{pmatrix}
    \alpha_\text{in}\\
    \beta_\text{in}
    \end{pmatrix}=\begin{pmatrix}
    \alpha_\text{out}\\
    \beta_\text{out}
    \end{pmatrix}
\end{equation}
where the intersections impose the relations
\begin{equation}
    U\begin{pmatrix}
    \alpha_\text{in}\\
    \phi_2
    \end{pmatrix}= \begin{pmatrix}
    \beta_\text{out}\\
    \phi_1e^{-i\xi_1}
    \end{pmatrix}\text{  and  } U\begin{pmatrix}
    \beta_\text{in}\\
    \phi_1
    \end{pmatrix}= \begin{pmatrix}
    \alpha_\text{out}\\
    \phi_2e^{-i\xi_2}
    \end{pmatrix}.
\end{equation}
After explicit calculation we arrive at 
\begin{subequations}
\begin{align}
    W&= \frac{1}{(1-P)e^{i(\xi_1+\xi_2+2\widetilde\varphi_S)}-1} \begin{pmatrix}
    Pe^{i\xi_1} & \kappa \\
    \kappa & Pe^{i\xi_2}\\
    \end{pmatrix}\\
    \kappa &= e^{-i\widetilde\varphi_S}\sqrt{1-P}(e^{i(\xi_1+\xi_2+2\widetilde\varphi_S)}-1),
\end{align}
\end{subequations}
Note that the basis we have chosen implies that when $W$ is diagonal, the amplitudes in different unit cells are decoupled. This occurs when $P=1$ (total magnetic breakdown) or when $\xi_1+\xi_2 + 2\widetilde\varphi_S \in 2\pi \mathbb{Z}$. The latter condition, combined with the quantization condition $\xi_1+\xi_2+\chi_1+\chi_2 \in 2\pi \mathbb{Z}$ for the full phase winding around the Fermi surface, implies a perfectly flat Chern band in the semiclassical model. \par 

The flat band wavefunctions, however, are generally not homogeneous along the orbit as would be the case for bona fide Landau levels. Rather, the density accumulates along the lens orbit due to the constructive interference of multiple reflections, suggesting a quantum analog of a Fabry-P\'erot optical cavity bound state. This is another way of seeing that the Hilbert space spanned by a magic zero is distinct from the free electron Landau level subspace. \par 
In the limit $P\to0$, $W$ is entirely off-diagonal with phase $i$, which is just the expected Maslov correction for turning points \cite{Xiao2010Jul,Keller1958Jun,Mikitik1999Mar,Fuchs2010Oct}. In this limit the network decouples into open orbits along the top and bottom and closed lens orbits. 

\textit{Density of states, 1D network. } There is a standard relation between a scattering matrix and its associated transfer matrix. We employ this relation, modified to include the phases $\chi_1,\chi_2$ on the top and bottom links, to obtain the transfer matrix
\begin{equation}
    T =  
    \frac{1}{W_{21}}
    \begin{pmatrix}
    -\det W  & W_{11}\\
    -W_{22} & 1\\
    \end{pmatrix} 
    \begin{pmatrix}
    e^{i\chi_1}&0\\0&e^{-i\chi_2}\\\end{pmatrix}.
\end{equation}
This describes evolution across one unit cell of the 1D network. Since $W_{21}=0$ when orbits decouple, it is already evident that there is a singularity at the magic condition. Physical states are required by the periodicity of $\p$-space to be extended eigenstates of $T$, i.e. states with pure phase eigenvalues $e^{i\theta}$ where $\theta$ can be considered a pseudomomentum. After a choice of gauge is made and the phases are determined as a function of $E_F$ and $B$, we can explicitly solve for the allowed bands. In the main text, we have chosen Landau gauge so that $\xi_1=\xi_2=\xi$ and $\chi_1=\chi_2=\chi$. It follows that
\begin{subequations}
\begin{align}
\chi &= l_B^2 (S_0-S_1)/2\\
\xi &= l_B^2S_1/2+\pi/2
\end{align}
\end{subequations}
where we the areas $S_0$ and $S_1$ and $\widetilde\varphi_S$ are defined in the main text. With this choice the determinant of $T$ is unity and the eigenvalues satisfy 
\begin{equation}\label{eq:costheta}
    \cos \theta = \frac12\Tr T.
\end{equation}
For parabolic dispersion, we make use of
\begin{equation}
    P = e^{-\pi \Delta^2/4E_F \omega_c \sin \alpha}
\end{equation}
where $\alpha = 2\sin^{-1}x$ and $x=q/2k_F$. The DOS can be explicitly obtained as $\rho(E_F,B) \propto |d\theta/dB|$. This model is closely related to the model considered by Pippard \cite{Pippard1962Oct}; we have extended it by applying a the unitary $U$ derived from first-principles (the precise form of which is important for recovering the perturbative limit), generalizing to arbitrary Fermi surface shape, extending to networks with many intersecting Fermi surfaces, including a Maslov correction to the spectrum, and extending, at least in part, to 2D networks at rational flux. \par
\begin{figure}
    \centering
 \includegraphics[width=1.0\columnwidth]{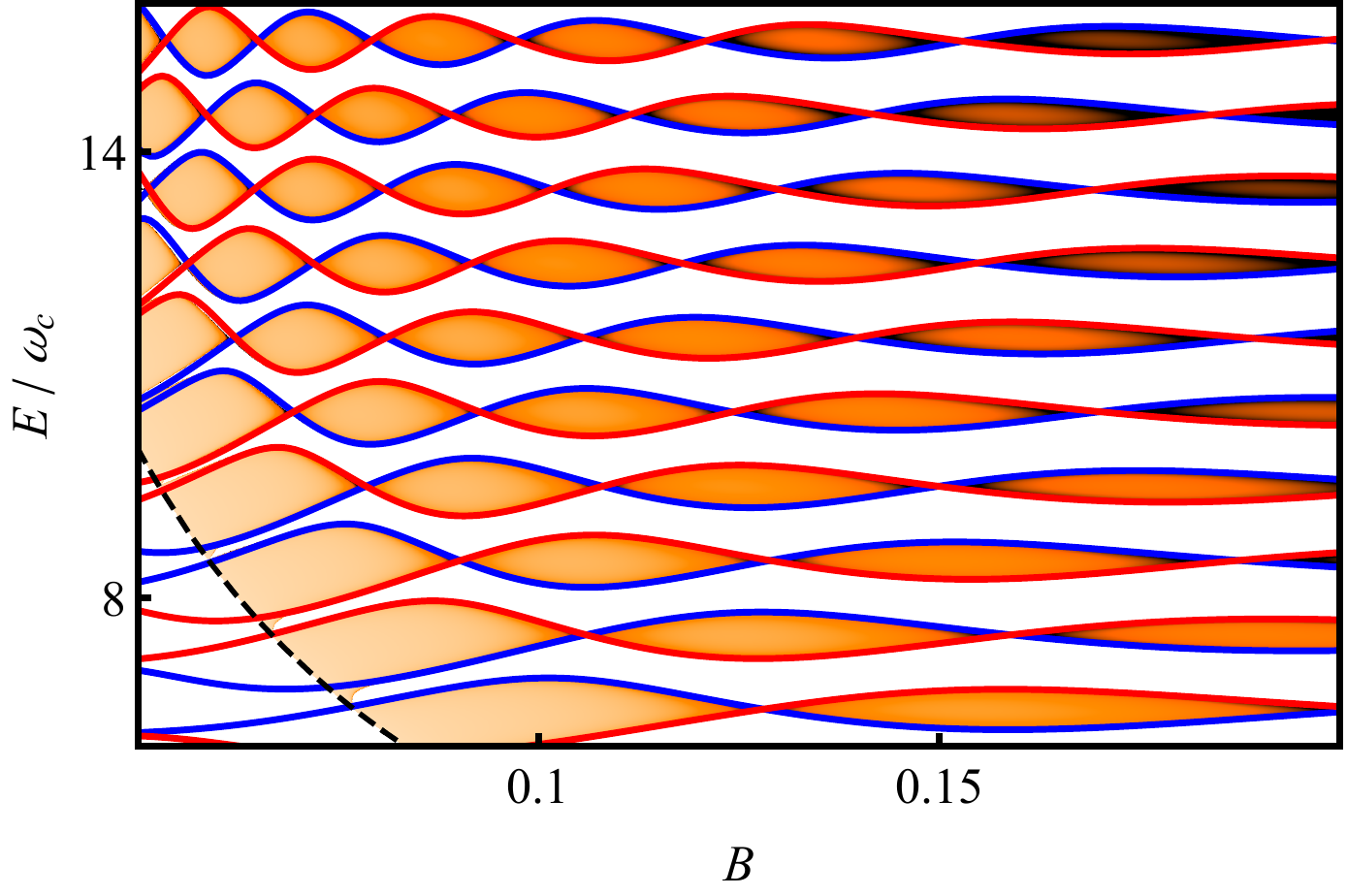}
    \caption{Comparing the magnetic spectrum obtained using semiclassics (density plot) and exact diagonalization (red and blue lines, showing band tops and bottoms). Plotted for parabolic dispersion $p^2/2m$ in a periodic potential $V_0 \cos qx$ with $V_0 =0.2,q=2,m=1$. For $E < q^2/8m$ (black dashed line) the orbits do not intersect and the semi-classical theory presented does not apply.}
    \label{fig:comp}
\end{figure}

 
In Fig. \ref{fig:comp}, we compare the DOS obtained using this semiclassical method with exact diagonalization (ED). Below a certain energy, shown as a dashed line, the semiclassical network model consists of non-intersecting Fermi surfaces and no band broadening is predicted, in contrast with the ED. However, the semiclassical approach is only expected to be valid when the Fermi energy is much larger than the potential strength, and in this regime we find good agreement. 

\begin{figure}
    \centering
    \includegraphics[width=1.0\linewidth]{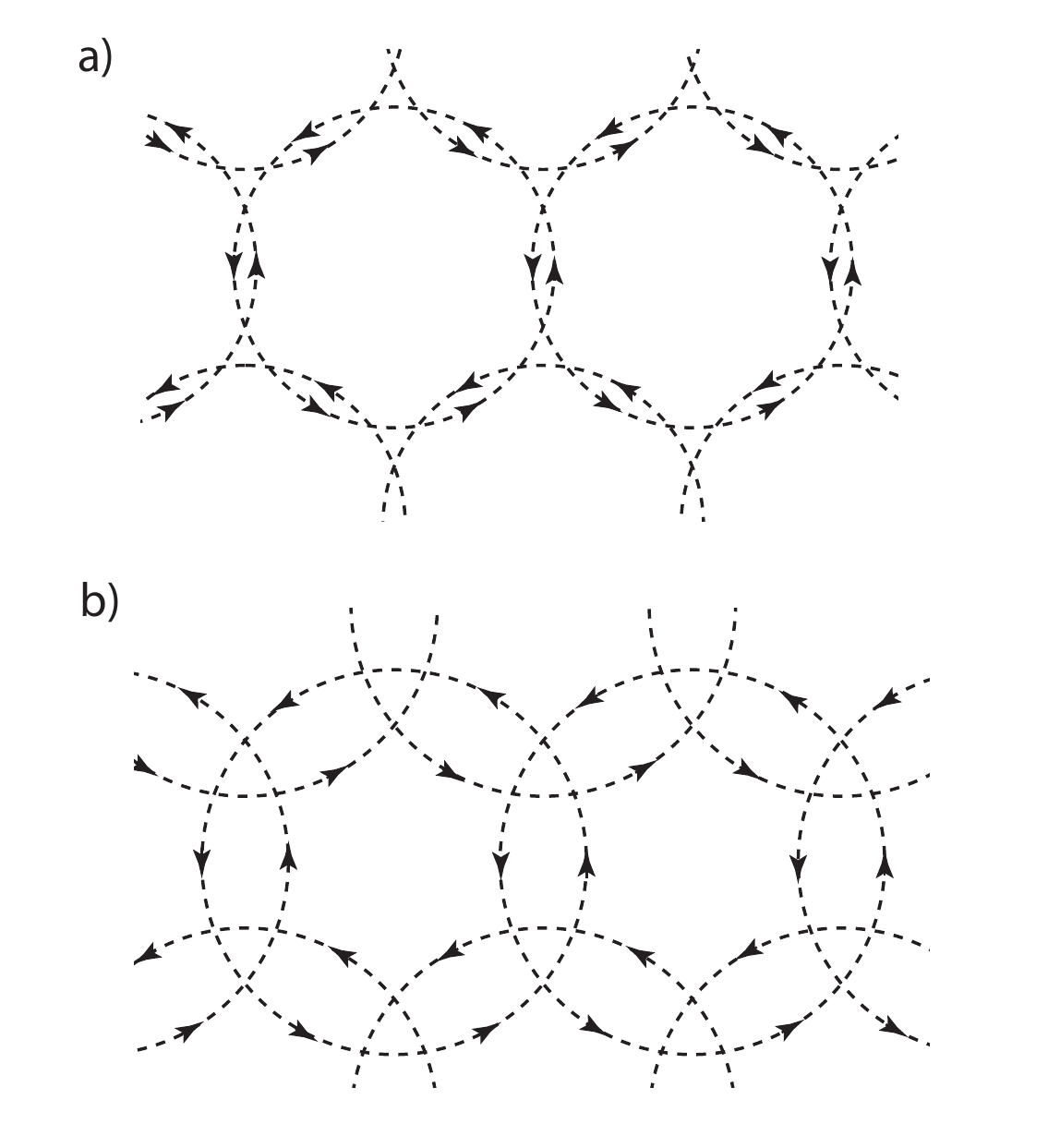}
    \caption{Network model schematic in the case of a potential with six-fold symmetry when (a) lens orbits do not overlap and (b) lens orbits overlap.}
    \label{fig:appendix2}
\end{figure}

\textit{Density of states, 2D hexagonal network. } We consider a periodic potential with six-fold symmetry and separate the discussion into two cases, shown in in Fig. \ref{fig:appendix2}. When the network topology takes the form of Fig. \ref{fig:appendix2}a, the arguments in the main text apply and there are flat bands at the magic zero conditions, with the following caveat: the network model itself has a well-defined unit cell (in $\p$-space) only when the number of flux quanta per unit cell of real space is a rational number $p/q$. However, the flux at any magic zero can be approximated arbitrarily well by a rational, so this does not pose any problems. This discussion mirrors the discussion in the main text. The second case is when the network topology takes the form of Fig. \ref{fig:appendix2}b, which is relevant when the Fermi energy exceeds $E_K$, the energy at the $K$ point of the mini BZ. The network has two pseudomomenta which we call $\theta_x, \theta_y$. When the number of flux quanta per unit cell is of the form $1/(2q)$ for integer $q$, the dispersion is captured by the following analog of Eq. \eqref{eq:costheta}:
\begin{widetext}
\begin{equation}\label{eq:C}
   C\equiv \sum_{i=1}^3\cos \vartheta_i= \frac{\sin(\frac12\mathcal{O})-2(1-P)^{3/2} \sin(\mathcal{O}-\frac32 \mathcal{L})+(1-P)^3\sin(\frac{3}{2}\mathcal{O}-3\mathcal{L})-3P^2(1-P)\sin(\frac12 \mathcal{O}-\mathcal{L})}{2P\sqrt{1-P}[\sqrt{1-P}\sin(\frac12 \mathcal{O}-\mathcal{L})-\sin(\frac12 \mathcal{L})]}
\end{equation}
\end{widetext}
where $\vartheta_1=\theta_x, \vartheta_2 = (-\theta_x-\sqrt{3}\theta_y)/2, \vartheta_3 = (-\theta_x+\sqrt{3}\theta_y)/2$ and
\begin{subequations}
\begin{align}
\mathcal{O}&=l_B^2 S_0+2\pi\gamma ,\\ \mathcal{L}&=l_B^2 S_1+2\widetilde \varphi_S + 2\pi\gamma,
\end{align}
\end{subequations}
are the original orbit and lens orbit Aharonov-Bohm phases, corrected by the Stokes phase and Maslov contributions. This result is adapted from Eq. 35 of Ref. \cite{Pippard1964Jul}, except that we've described scattering at junctions by the Landau-Zener unitary $U$ and Ref. \cite{Pippard1964Jul} lacked a first-principles calculation for this unitary. \par
We may note that $C$ becomes singular when both $\mathcal O$ and $\mathcal L$ are integer multiples of $2\pi$, and this is equivalent to the magic zero conditions in the main text. The bandwidth is zero whenever this condition is met. However, we are not currently aware of the generalization to all rational fluxes. When the number of flux quanta per unit cell is $p/(2q)$, the network model unit cell must be enlarged by a factor of $p$, and a difficult set of equations must be solved. However, it is reasonable to expect that the band envelopes of Eq. \eqref{eq:C} are quite accurate at all flux values, while the exact solution of the network model only refines the subband structure \cite{Chambers1965Oct}. While a proof of this is lacking, we note that there are still infinitely many magic zeros arbitrarily close to fluxes of the form $1/(2q)$; moreover, we proved in the main text that magic zeros are present when the network model consists of non-overlapping lens orbits. \par 
Finally, we note that six-fold (or four-fold) symmetry of the potential is important. An anisotropic potential does not even lead to perfectly flat bands in the perturbative regime. Equal magnitude wavevectors is the most common scenario in moir\'e materials, however.\par

 \textit{Hofstadter model. } While both the Landau spectrum in a 2D moiré potential and the Hofstadter spectrum exhibit fractal-like features, our study considers Landau levels in a weak superlattice potential relative to bandwidth while the conventional Hofstadter model considers a single band in the tight-binding limit of deep potential. Moreover, perfectly flat bands are not present in the conventional Hofstadter model: while some bands become flat as the flux approaches an integer or half-integer, these bands also have vanishing spectral weight. 

\textit{Weiss oscillations. } Weiss oscillations show up in conductivity and magnetoresistance measurements in addition to the usual Shubnikov-de Haas (SdH) oscillations when a periodic potential is present \cite{Weiss1989Jan}. The hallmark is that the DOS peaks when the cyclotron radius is commensurate with the period \cite{Gerhardts1989Mar,Zhang1990Jun,Steda1990Jun,Pfannkuche1992Nov,PhysRevLett.121.026806}: 
\begin{equation}\label{eq:2Rc}
    2R_c = a(j-1/4),\qquad j=1,2,3....
\end{equation}
where $R_c = k_F/e B$ for circular Fermi surfaces. It is simple to see that this is identical to the large $n$, perturbative result in the main text, which is derived within the semiclassical framework. The semiclassical description of the magic zeros as the intersection of two Landau fans, however, does not rely on any particular dispersion or on the perturbative limit. \par 
Any SdH oscillation can be written as a quantization condition $l_B^2 S = 2\pi (j+\phi)$ where $S$ is some $\p$-space area. Naturally, we may ask if there is an interpretation for Weiss oscillations in terms of a $\p$-space area, using the fact that Eq. \eqref{eq:2Rc} can be written 
\begin{equation}
    l_B^2 (4k_F^2x) = 2\pi (j-1/4)
\end{equation}
where $x=q/2k_F$. Indeed, it follows that
\begin{equation}
    4k_F^2x = S_0 - S_1 + O(n^{-\frac12})
\end{equation}
(using $k_F \sim \sqrt{n}$) so in the large $n$ limit our semiclassical model associates Weiss oscillations to the $\p$-space area $S_0-S_1$. This is not the area of any semiclassical orbit, as would be the case for SdH oscillations, but rather the \textit{difference} of such areas. A suggestive interpretation of the above is as follows. The magic zeros occur at the intersections of two Landau fans corresponding to oscillation frequencies $S_0$ and $S_1$, respectively. Roughly speaking, the frequency of intersections of the two fans is then the difference of frequencies. Indeed, the observation of quantum oscillations which cannot be associated to any semiclassical orbit is expected when magnetic breakdown is taken into account \cite{Pippard1962Oct}.

\section{Application to Twisted Bilayer Graphene}

 \begin{figure}
    \centering
    \includegraphics[width=1.0\columnwidth]{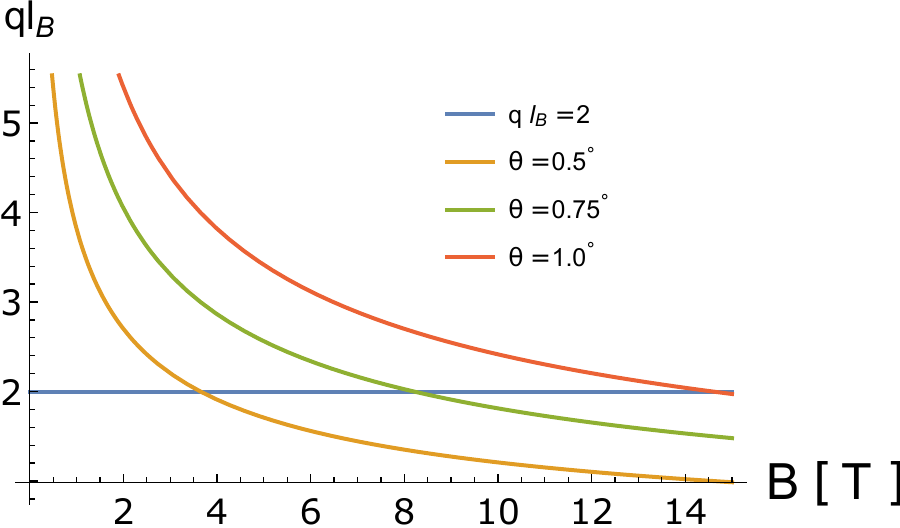}
    \caption{Plots of $q l_B$ for various twist angles (colored lines) in the chiral limit of TBG  as well as the magic zeros in the 2nd and 3rd LLs. Various magic zeros are accessible at realistic fields.}
    \label{fig:tbg}
\end{figure}

In this section we investigate whether magic zeros arise in twisted bilayer graphene (TBG) in a uniform magnetic field. The spectrum of TBG is described by \cite{Koshino2018Sep}

\begin{equation}
    H = v (\vec k \cdot \vec \sigma) \mathbb{I}+ \begin{pmatrix}0&U(r)\\U(r)^\dagger&0 \end{pmatrix}
\end{equation}
where $U(r) = \sum_{j=1}^3 U_n e^{i \q_j \cdot \r}$ where $\q_1 = q (0,1)$ and $\q_{2,3} = q (\pm \sqrt{3}/2,-1/2)$ and 
\begin{equation}
    U_{j+1} = w_{AA} \sigma_0 + w_{AB} (\sigma_x \cos \frac{2\pi j}{3} + \sigma_y \sin \frac{2 \pi j}{3}).
\end{equation}
The Dirac Landau levels when $U=0$ are 
\begin{equation}
    \ket{n,\xi}_D = \frac{1}{\sqrt{2}} \begin{pmatrix}
    \ket{n}\\
    \ket{n-1}\end{pmatrix}\otimes \ket{\xi}_L
\end{equation}
where $\ket{\xi}_L$ $(\xi=\pm 1)$ is a layer eigenstate and $\ket{n}$ is a Schr\"odinger LL wavefunction. We've assume $n>0$ for now. The potential projected into the $n$'th Dirac LL is
\begin{equation}\begin{pmatrix} 0 & \braket{n,1|U|n,2}_D\\ \braket{n,2|U^\dagger |n,1}_D & 0\end{pmatrix}
\end{equation}
where \cite{Wilkinson1987May}
\begin{widetext}
\begin{multline}
    \braket{n,1|U|n,2}_D = \frac12 w_{AA} e^{-q^2l_B^2/4}(L_n(q^2l_B^2/2) + L_{n-1}(q^2l_B^2/2)) \sum_{j=1}^3 e^{i\q_j \cdot \hat \r}\\
    +\frac{1}{\sqrt{2}}w_{AB}  e^{-q^2l_B^2/4}ql_BL^{(1)}_{n-1}(q^2l_B^2/2) \sum_{j=1}^3 \cos \frac{2\pi(j-1)}{3} e^{i2\pi (j-1)/3} e^{i\q_j\cdot \hat \r}
    \end{multline}
\end{widetext}
where $\hat \r = (\hat X, -\hat P)$ and $[\hat X, \hat P] = il_B^2$. This can be extended to all levels by replacing $n$ with $|n|$, with the convention $L_{-1}(x) = L_{-1}^{(1)}(x) = 0$. Let's take the chiral limit, $w_{AA}=0$. Then it is clear that the projected potential vanishes whenever 
\begin{equation}
    L_{n-1}^{(1)}(q^2l_B^2/2) = 0.
\end{equation}
In the anti-chiral limit of $w_{AB}=0$, the condition is 
\begin{equation}
    L_n(q^2l_B^2/2) + L_{n-1}(q^2l_B^2/2) = 0.
\end{equation}
 Examples of magic zero values for both limits are provided in Table \ref{tbgtable}. In realistic TBG, it is possible that magic zeros corresponding to either of these two limits are broadened but still observable (see e.g. Fig. 1 of \cite{Herzog-Arbeitman2021Nov}). \par

We note that $q = \frac{8\pi}{3a_0} \sin\frac{\theta}{2}$ at small angles in TBG, where $a_0 = 0.246$ nm is the lattice constant of graphene. A useful relation is then $q l_B = 875.2 \sin\frac{\theta}{2}/\sqrt{B\text{[T]}}.$ In Fig. \ref{fig:tbg} we plot this relation for various $\theta$ and compare to a few magic zeros. This indicates the ``approximate magic zeros" are accessible at realistic magnetic fields.

\begin{table}[h!]
\centering
\begin{tabular}{l|l|l}
\textrm{$|n|$}&
$ql_B$ \textrm{(chiral limit)} & $ql_B$ \textrm{(anti-chiral limit)}\\
\hline
$1$ & & 2\\
$2$ & 2 & 1.24, 3.24 \\
$3$ & 1.59, 3.08 & 0.99, 2.36, 4.18\\
$4$ & 1.37, 2.57, 3.94 & 0.86, 2\,, 3.26, 4.96 \\
$5$ &1.22, 2.27, 3.39, 4.68 & 0.76, 1.77, 2.84, 4.03, 5.65  \\
\end{tabular}
\caption{ Magic zero values of $ql_B$ for the $n$th Dirac Landau level in TBG, in the chiral and anti-chiral limits, at first order in perturbation theory.}\label{tbgtable}
\end{table}

\section{Compressibility signatures} 

\begin{figure}[t]
    \centering
    \includegraphics[width=1.0\columnwidth]{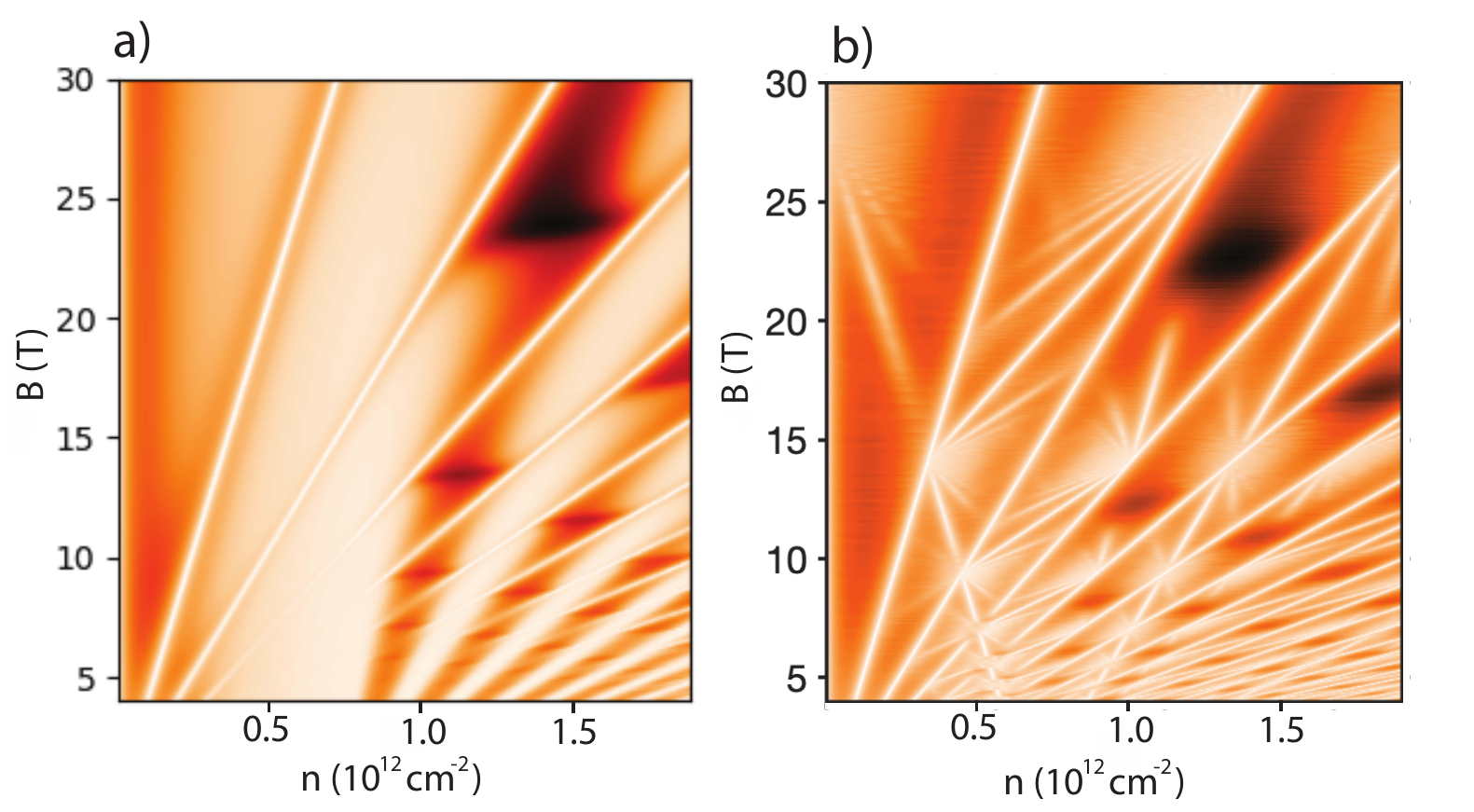}
    \caption{  Noninteracting compressibility as a function of density and $B$ for a 1D potential (a) and 2D potential (b), at $T = 0.2$ meV. Magic zeros are dark features of high compressiblity occurring over a finite range of $n$. Panel (b) exhibits Hofstadter-like features in addition to the magic zeros. Parameters match Fig. 1a and 1c, respectively.}
    \label{fig:compressibilitySM}
\end{figure}

We include here some comments on the compressibility signatures of magic zeros. The flat bands due to magic zeros would na\"ively lead to strong compressibility signatures. In Fig. \ref{fig:compressibilitySM} we plot the compressibility due to the magnetic spectrum in 1D and 2D moir\'e potential cases. \par
In real systems, however, interactions would destabilize these sharp features and possibly lead to interesting correlated states near the magic zeros. The nature of the correlated states near magic zeros is an interesting problem for future work.

\bibliographystyle{apsrev4-1}
\bibliography{bib}